%% file: main.tex
\documentclass[aps,twocolumn,prd,showpacs,showkeys,preprintnumbers,superscriptaddress,nobibnotes,floatfix,longbibliography,nofootinbib]{revtex4-1}

\pdfoutput=1
\include{jdefs}
\usepackage{tikz}
\usepackage[compat=1.1.0]{tikz-feynman}
\usepackage{feynmp-auto}
\usetikzlibrary{calc}
\usetikzlibrary{shapes.misc}
\tikzset{
	>=latex,
    photon/.style={decorate, decoration={snake}, draw=black, thick},
    fermionnoarrow/.style={draw=black, postaction={decorate}, thick},
    scalar/.style={draw=black, postaction={decorate}, decoration={markings,mark=at position .55 with {\arrow{>}}}, thick, dashed},
    scalarnoarrow/.style={draw=black, postaction={decorate},  thick, dashed},
    fermion/.style={draw=black, postaction={decorate},decoration={markings,mark=at position .55 with {\arrow{>}}}, thick},
    gluon/.style={decorate, draw=black, decoration={coil,amplitude=4pt, segment length=5pt}, thick},
    vertex/.style={draw,shape=circle,fill=black,minimum size=3pt,inner sep=0pt},
    fillvertex/.style={draw,shape=circle,fill=black,minimum size=5pt,inner sep=0pt},
    openvertex/.style={draw,shape=circle,minimum size=51pt,inner sep=0pt},
    blob/.style={draw=black,shape=circle,fill=black,minimum size=6pt,inner sep=0pt},
    redvertex/.style={draw=red,shape=circle,fill=red,minimum size=3pt,inner sep=0pt},
    cross/.style={cross out, draw=black,thick, minimum size=5pt, inner sep=0pt, outer sep=0pt}
}
\usepackage{amsmath}
\usepackage{amsfonts}
\usepackage{amssymb}
\usepackage{mathrsfs}
\usepackage{graphicx}
\usepackage{xcolor}
\usepackage{longtable}
\usepackage[colorlinks]{hyperref}
\hypersetup{
    colorlinks,
    linkcolor={magenta},
    citecolor={magenta},
    urlcolor={magenta}
}
\usepackage{multirow}
\usepackage{slashed}

\usepackage{array}

\definecolor{lime}{HTML}{A6CE39}
\DeclareRobustCommand{\orcidicon}{
	\begin{tikzpicture}
	\draw[lime, fill=lime] (0,0) 
	circle [radius=0.16] 
	node[white] {{\fontfamily{qag}\selectfont \tiny ID}};
	\draw[white, fill=white] (-0.0625,0.095) 
	circle [radius=0.007];
	\end{tikzpicture}
	\hspace{-2mm}
}

\foreach \x in {A, ..., Z}{\expandafter\xdef\csname orcid\x\endcsname{\noexpand\href{https://orcid.org/\csname orcidauthor\x\endcsname}
			{\noexpand\orcidicon}}
}

\begin{document}

\preprint{
\begin{minipage}{5cm}
\small
\flushright
UCI-HEP-TR-2024-17
IFT-UAM/CSIC-24-156
\end{minipage}} 

\title{Boxed in from all sides: a global fit to loopy dark matter and neutrino masses}

\author{Karen Mac\'ias C\'ardenas\orcidA{}}
\thanks{\href{mailto:karen.macias@csic.es}{karen.macias@csic.es}, she/her and they/them.}
\affiliation{Instituto de F\'isica Te\'orica, IFT-UAM/CSIC, 28049 Madrid, Spain}
\affiliation{Department of Physics, Enigneering Physics and Astronomy, Queen's University, Kingston, ON, K7L 3N6, Canada}
\affiliation{Arthur B. McDonald Canadian Astroparticle Physics Research  Kingston ON K7L 3N6, Canada}

\author{Gopolang Mohlabeng\orcidB{}}
\thanks{\href{mailto:gmohlabe@sfu.ca}{gmohlabe@sfu.ca}, he/him and all pronouns.}
\affiliation{Department of Physics, Simon Fraser University, Burnaby, BC, V5A 1S6, Canada}
\affiliation{TRIUMF, 4004, Westbrook Mall, Vancouver, BC, V6T 2A3, Canada}
\affiliation{Department of Physics and Astronomy, University of California, Irvine, CA 92697, USA}

\author{Aaron C. Vincent\orcidC{}}
\thanks{\href{mailto:aaron.vincent@queensu.ca}{aaron.vincent@queensu.ca}, he/him.}
\affiliation{Department of Physics, Enigneering Physics and Astronomy, Queen's University, Kingston, ON, K7L 3N6, Canada}
\affiliation{Arthur B. McDonald Canadian Astroparticle Physics Research  Kingston ON K7L 3N6, Canada}
\affiliation{Perimeter Institute for Theoretical Physics, Waterloo ON N2L 2Y5, Canada}

\date{\today}

\begin{abstract}
We investigate a dark matter model that couples to the standard model through a one-loop interaction with neutrinos, where the mediator particles also generate neutrino masses. We perform a global fit that incorporates dark matter relic abundance, primordial nucleosynthesis, neutrino mass, collider and indirect detection constraints. Thanks to the loop suppression, large couplings are allowed, and we find that the model parameters are constrained on all sides. Dark matter masses from 10 MeV to a few TeV are allowed, but sub-GeV masses are preferred for the model to also account for the heaviest neutrino mass. Though our results are valid for a single neutrino mass eigenstate at a time, the model and methods are generalizable to the full 3-flavor case.

\end{abstract}

\maketitle
\section{Introduction \label{sec:intro}}

The existence of cosmic dark matter (DM) and non-zero neutrino masses and mixing are arguably two of the most compelling pieces of evidence for new physics beyond the Standard Model (SM) \cite{Bertone:2016nfn, PDG2018, Super-Kamiokande:1998kpq, SNO:2003bmh}.
However, despite years of experimental and theoretical investigations, the fundamental nature of dark matter and the mechanism through which neutrinos obtain their masses remain unknown. Since the interaction strength of both the neutrino and dark matter sectors lead to vanishingly small cross sections with the SM, it is natural to ask about their possible connections, leaving room for contemplation about the strength of their interaction.\\

This provides exciting opportunities for understanding the underlying properties of both dark matter and neutrinos. 
However, it presents a challenge to probe given the feeble interactions of both sectors with the rest of the SM.
Still, significant work has gone into investigating such a possibility.
A large coupling between dark matter and neutrinos can leave imprints on cosmological observables such as the CMB and large scale structure
\cite{Boehm:2000gq, Boehm:2004th, Mangano:2006mp, serra2009constraints, vandenAarssen:2012vpm, Wilkinson:2014ksa, Schewtschenko:2014fca, Boehm:2014vja, Escudero:2015yka, Olivares-DelCampo:2017feq, OttaviaBalducci:2017abc, Wilkinson:2014ksa, Bertoni:2014mva, DiValentino:2017oaw, Diacoumis:2018ezi, Mangano:2006mp, Serra:2009uu,Hooper:2021rjc}.
It may also manifest itself in galaxies \cite{Akita:2023yga, Cline:2023tkp} as well as other astrophysical processes. 
For instance, ultra high energy neutrinos of astrophysical origin can scatter with ambient dark matter and lead to attenuation of the expected neutrino spectrum at Earth \cite{Reynoso:2016hjr, Arguelles:2017atb, Vincent:2017svp, Arguelles:2019rbn, Choi:2019ixb}.
Moreover, lower energy neutrinos, originating from stars, supernovae or the cosmic neutrino background can scatter with ambient dark matter, resulting in both a modification of the expected spectrum at Earth \cite{Farzan:2014gza, Farzan:2018pnk, Carpio:2022sml, Balantekin:2023jlg} as well as up-scattering of dark matter \cite{Zhang:2020nis, Das:2021lcr, Jho:2021rmn, Ghosh:2021vkt}.
These searches are complemented by model-independent investigations of galactic and extra-galactic dark matter annihilating or decaying directly into neutrinos (see refs.~\cite{Arguelles:2019ouk, Buckley:2022btu, Arguelles:2022nbl, Das:2023wtk, Fiorillo:2023clw, Song:2023xdk,Allahverdi:2023nov,Fuss:2024dam, Das:2024bed} and references therein).

On the theory side, the landscape of scenarios that provide an explanation for dark matter-neutrino interactions is broad.
Many of the underlying models provide a viable candidate for dark matter while explaining the origin of neutrino masses, and provide complementary signals in terrestrial experiments.
Some common examples include the scotogenic and scotogenic-like models (see ref.~\cite{Ma:1998dn,Ma:2006km,Boehm:2006mi},  more recently \cite{Herms:2023cyy,Jana:2024iig}, and \cite{Borah2019} for a connection to leptogenesis), and the Zee model (see ref.~\cite{Zee:1980ai} and references thereafter). 
These scenarios are well-motivated, but come with many parameters, making them hard to constrain. It is common in the literature to fix some parameters to ad hoc benchmark values while varying others to show the viable parameter space. This can lead to misinterpretations of the parameter space, depending on the parameters represented and their constraints.

In this work, we explore a scenario in which the leading coupling of the dark matter to the neutrinos occurs at 1-loop level. 
The scenario is motivated by scotogenic models. 
In contrast with scotogenic models, the dark matter production is loop-suppressed, allowing thermal production through large couplings, while the unstable mediator particles are responsible for neutrino mass generation. 

To explore the rich phenomenology presented by this model, we perform a global likelihood analysis and investigate the terrestrial and astrophysical constraints on the multi-dimensional parameter space, without fixing any parameters. Remarkably, we will find that for couplings that are small enough for the theory to remain perturbative, constraints from cosmology, indirect detection, neutrino oscillation, and collider experiments bound this model to a closed region of parameter space. This model naturally points to an MeV--GeV mass dark matter candidate $\eta$. While this model may contain a second region of parameter space at lower couplings leading to dark matter production via freeze-in (as opposed to freeze-out), this would likely obviate some of the benefits mentioned above. We leave this technically distinct exploration for future work.

The rest of this article is organized as follows: In section \ref{sec:model} we describe the model and the relevant parameters. In section \ref{sec:pheno}, we describe the observables and experimental constraints, and in section \ref{sec:results} we show the results of our global fit. We conclude in section \ref{sec:discussion}.

\section{Model}\label{sec:model}
We consider a scenario in which a real scalar dark matter particle $\eta$ has leading interactions with the SM at 1-loop level, through a Majorana fermion $N_R$ and a complex scalar field $\Phi$.
The Lagrangian contains the interaction terms
\begin{equation}
\begin{aligned}
    \mathcal{L}_{\mathrm{int}} = & - (Y_{\nu } \overline{\nu}{\Phi}{N_R}+ \mathrm{h.c.})-m_\Phi^2|\Phi|^2-\frac{1}{2}\mu^2(\Phi^2 + (\Phi^\dagger)^2) \\& - \frac{\lambda^\prime}{2}\eta^2|\Phi|^2 - \frac{\lambda}{2}\eta^2\left(\frac{\Phi^2+(\Phi^\dagger)^2}{2}\right),
\end{aligned}
\label{eq:lagrangian}
\end{equation}
where $\lambda^{\prime}$ and $\lambda$ are the quartic couplings linking the DM to the complex scalar mediator, $Y_{\nu}$ is a Yukawa-type coupling linking the dark sector to the left-handed neutrinos of the SM. $\mu$ is a constant that splits the U(1) symmetry of $\Phi$ as in the SLIM mechanism of Ref. \cite{Boehm:2006mi}, ultimately leading to the neutrino mass terms. Interactions between the SM Higgs and the new scalars, as well as the quartic coupling $\lambda^\prime$, are neglected for simplicity as in Ref. \cite{Chao:2020bti}.

We decompose $\Phi$ into its real and imaginary parts, $\Phi = (\phi_1+i\phi_2)/\sqrt{2}$, after which:
\begin{equation}
\begin{aligned}
    \mathcal{L}_{\mathrm int} = & - (Y_{\nu } \overline{\nu}{\Phi}{N_R}+ \mathrm{h.c.})-\frac{1}{2}m_{\phi_1}^2\phi_1^2-\frac{1}{2}m_{\phi_2}^2\phi_2^2 \\& - \frac{\lambda^\prime}{4}\eta^2(\phi_1^2+\phi_2^2) - \frac{\lambda}{4}\eta^2(\phi_1^2-\phi_2^2).
    \label{eq:effective_lagrangian}
\end{aligned}
\end{equation} 
The quantities $m^2_{\phi_{1,2}} = m_{\Phi}^{2} \pm \mu^{2}$ are the masses of $\phi_{1,2}$ respectively, and by construction,  $m_{\phi_1}>m_{\phi_2}$.
This phenomenology can be straightforwardly generalized to three flavors. However, for simplicity, we focus on a single mass eigenstate at a time.

We choose this model for two main reasons. First, with an appropriately chosen hierarchy of masses, the dark matter annihilation rate becomes loop-suppressed, leading to a natural mechanism for obtaining the weak cross section required for thermal freeze-out dark matter production. Second, the mass splitting introduced above can generate neutrino masses through the scotogenic-like mechanism described in \cite{Boehm:2006mi}: 
\begin{equation}
\begin{aligned}
m_{\nu}=\frac{Y_\nu^{2}}{32 \pi^{2}} m_{N} & {\left[\frac{m_{\phi_ 1}^{2}}{\left(m_{N}^{2}-m_{\phi_1}^{2}\right)} \ln \left(\frac{m_{N}^{2}} { m_{\phi_1}^{2}}\right)\right.} \\&\left.-\frac{m_{\phi_2}^{2}}{\left(m_{N}^{2}-m_{\phi_2}^{2}\right)} \ln \left(\frac{m_{N}^{2} }{ m_{\phi_2}^{2}}\right)\right].
\end{aligned}
\label{eq:neutrino_mass}
\end{equation}
The direct implication is that neutrino masses are generated by a Lepton Number Violating (LNV) process. Setting $m_{\phi_1}=m_{\phi_2}$ restores lepton number conservation, and neutrino masses vanish. Going forward, we parametrize this with the dimensionless quantity 
\begin{equation}
\delta m_\phi = 1 - \frac{m_{\phi_2}}{m_{\phi_1}},
\label{eq:deltamphi}
\end{equation}
where $\delta m_\phi = 0$ implies the neutrino is massless. 

As noted in Refs. \cite{ErnestMa, Gonzalez-Macias:2016vxy, Rodejohann:2019quz, Chao:2020bti}, imposing a $Z_2$ symmetry on the real scalar dark matter candidate $\eta$ forbids direct couplings to the neutrinos. Furthermore, this symmetry forbids tree-level contributions to the left-handed neutrino masses as in the classic seesaw scenario. In our model, the source of the LNV process is the Majorana mass of $N$.

\begin{figure}[!ht]
    \centering
        \includegraphics[width=0.49\columnwidth]{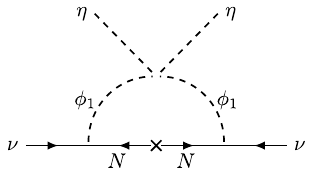}
    \includegraphics[width=0.49\columnwidth]{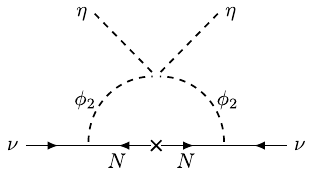}
    \includegraphics[width=0.49\columnwidth]{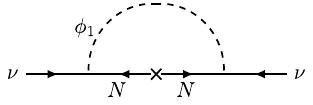}
    \includegraphics[width=0.49\columnwidth]{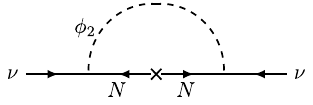}

    \caption{Top: loop processes leading to dark matter-neutrino interaction in the model of Eq.~\eqref{eq:effective_lagrangian}, allowing annihilation and elastic scattering. Bottom: diagrams contributing to the neutrino mass generation.}
    \label{fig:feynman_diagrams}
\end{figure}

In the absence of tree-level processes, Fig. \ref{fig:feynman_diagrams} shows the leading diagrams that will govern dark matter-neutrino interactions, generated with {\fontfamily{lmtt}\selectfont  FeynArts} \cite{Feynarts_Hahn_2001} using the Lagrangian in Eq. \eqref{eq:effective_lagrangian}. These are similar to the well-known scotogenic model (see \cite{Ma:2006km,Cai:2017jrq}), with a few key differences.  In the scotogenic model, the scalar mediator is the dark matter particle, and it interacts at tree level with the SM Higgs, gaining its mass in that way. In contrast, we introduce the scalar field $\eta$, which is stabilized by a $Z_2$ symmetry and does not acquire a vacuum expectation value ($vev$).

In this picture, a new possibility also appears compared to the scotogenic model: in the hierarchy $m_N>m_\Phi>m_\eta$, the main dark matter candidate is $\eta$. This means that, even though $\Phi$ and $N$ could still be dark matter candidates, the main channel to produce dark matter is not through the tree-level contributions of the mediators annihilating to neutrinos as in the scotogenic case, but through $\eta$ annihilating into neutrinos at the one-loop level. We implicitly assume this mass hierarchy for the rest of this work.

\section{Phenomenological analysis}
\label{sec:pheno}

Despite its relative simplicity, this model offers a wealth of phenomenological predictions, and ends up being viable in a closed region of parameter space. First, the two-to-two annihilation of $\eta$ particles into neutrinos offers a dark matter generation mechanism via thermal freeze-out. The same process also leads to late-time annihilation, which can be constrained by the (non) detection of a neutrino line flux at neutrino telescopes. By crossing symmetry, the same diagrams lead to elastic scattering between dark matter and neutrinos, which is constrained by the Cosmic Microwave Background (CMB) and small-scale structure like the Lyman-$\alpha$ forest, as well as the Milky Way satellites.  
If the dark matter is lighter than a few MeV, strong constraints exist due to its contribution to the effective number of ultrarelativistic degrees of freedom ($N_\text{eff}$) during Big Bang Nucleosynthesis (BBN). Additionally, new interactions with neutrinos will lead to new final-state kinematics in meson and Z boson decays. Finally, the upper limit on the lightest neutrino mass constrains the couplings and masses that generate the lightest neutrino mass, while oscillation data further restricts the mass of the heaviest neutrino. We will examine the cases of lightest and heaviest neutrino separately.\\

In the mass basis, and considering a single neutrino eigenstate, this model has 6 free parameters: $\theta=\{m_\eta,m_{\phi_1},\delta m_\phi, m_N, \lambda, Y_\nu\}$. Here $m_\eta$ is the dark matter mass, $m_{\phi_1}$ is the mass of the real scalar mediator, $\delta m_\phi$
 is the mass splitting, Eq.~\eqref{eq:deltamphi}, between the scalar and pseudoscalar parts of the complex mediator $\Phi$ and is proportional to the neutrino mass through Eq.~\eqref{eq:neutrino_mass}, $m_N$ is the mass of the neutral singlet mediator $N$, $\lambda$ is the coupling for the scalars and $Y_\nu$ is the Yukawa-like coupling between $N$, $\Phi$ and $\nu$ from the Lagrangian Eq.~\eqref{eq:lagrangian}. 

Given these, we will compute the observable rates detailed above, and test the model against data from cosmology, astrophysics and particle physics.
We perform scans of the parameter space using the Markov Chain Monte Carlo (MCMC) sampler {\fontfamily{lmtt}\selectfont 
emcee} \cite{Foreman_Mackey_2013} with  log-flat priors on each parameter, which are given in Table \ref{tab:priors}. As previously discussed, we impose the mass hierarchy $m_N>m_{\phi_1}>m_\eta$ and additionally the perturbativity of the couplings $\lambda \leq 4\pi$, $Y_\nu \leq \sqrt{4\pi}$ as priors.

\begin{table*}[]
\centering
\caption[Log-flat priors]{Description and log-flat priors for all the parameters of the model (Eq. \ref{eq:lagrangian}). A dash (—) entry in the units column corresponds to dimensionless quantities.}
\label{tab:priors}
\begin{tabular}{l l l l l}
\hline
Parameter       & Description &  Log-flat prior & Units \\ \hline
$m_\eta$        & Dark matter mass &  [-4,8]  & GeV     \\ 
$m_{\phi_1}$    & Scalar mediator mass &  [-4,8]  & GeV       \\ 
$\delta m_\phi$ & Mass splitting of the scalar and pseudoscalar mediators  &  [-16,0] & --    \\ 
$m_N$           & Neutral singlet mediator mass  &  [-4,8] & GeV      \\ 
$\lambda$       & Quartic coupling for the term $\frac{\lambda}{4}\eta^2(\phi_1^2-\phi_2^2)$ &  [-7,log$_{10}(4\pi)$] & --      \\ 
$Y_\nu$         & Yukawa-like coupling for the term $Y_{\nu } \overline{\nu}{\Phi}{N_R}$ + h.c. &  [-7,log$_{10}(\sqrt{4\pi})$]  & --        \\ \hline
\end{tabular}%
\end{table*}

We construct the following likelihood function:
\begin{equation}
\begin{aligned}
    -2 \ln \mathcal{L}_{\mathrm{total}} = & -2 \ln \mathcal{L}_{\mathrm{RA}} -2 \ln \mathcal{L}_{\mathrm{ann}} -2 \ln \mathcal{L}_{N_\mathrm{eff}} \\& -2 \ln \mathcal{L}_{\mathrm{sc}} -2 \ln \mathcal{L}_{m_{\nu_{1,3}}},
    -2 \ln \mathcal{L}_{\text{decays}},
\end{aligned}
    \label{eq:likelihood}
\end{equation}
where the terms on the right-hand side are, in order, the  dark matter relic abundance, the late-time annihilation rate to observable neutrinos, $N_\text{eff}$ at BBN, dark matter-neutrino scattering, the lightest or the heaviest neutrino mass, and finally Z boson and charged meson decays. We now turn to the definitions of these individual pieces. 

\subsection{Dark matter relic abundance}
The relic abundance of the stable dark matter candidate $\eta$ is obtained from the thermal freeze-out process. The dark matter annihilation cross-section is computed from the loop diagrams in Fig.~\ref{fig:feynman_diagrams}. We present the full procedure in App. \ref{app:crosssections}. The cross section can be written as
\begin{equation}
\begin{aligned}
    \langle \sigma v \rangle = \frac{\lambda^2 Y_\nu^4 m_N^2}{4096\pi^5} & \big|C_0(\phi_1)+C_0(\phi_2)\big|^2 \, ,
    \label{eq:svresult}
\end{aligned}
\end{equation}
 here we have defined $C_0(\phi_i) \equiv C_0(0,0,s,m_{\phi_i}^2,m_N^2,m_{\phi_i}^2)$, where $C_0(p_1^2, p_2^2,\left(p_1+p_2\right)^2, m_1^2, m_2^2, m_3^2)$ is the conventionally-defined scalar triangle integral (see e.g. \cite{looptoolsHahn_1999}). For the special case here, the squared matrix element in Eq.~\eqref{eq:svresult} can be written in terms of  logarithmic integrals, given in Eq.~\eqref{eq:C0expansion}.

We have constructed a fully numerical Boltzmann solver to track the relic abundance through freeze-out to the present day, including the full temperature dependence of $\langle\sigma v\rangle$, which we have benchmarked against the DRAKE \cite{Binder_2021} package. This method is too computationally intensive to be used in the full MCMC. We thus take only the $s-$wave ($\langle \sigma v \rangle = $ constant) contribution to the cross-section and employ an approximation to the relic abundance following the methods outlined in Ref. \cite{Steigman:2012nb}:
\begin{equation}
    \Omega_\eta h^{2}=\frac{9.92 \times 10^{-28} \mathrm{cm}^{3} \, \mathrm{s}^{-1} }{\langle\sigma v\rangle   } \left(\frac{x_{*}}{g_{*}^{1 / 2}}\right)\left(\frac{(\Gamma / H)_{*}}{1+\alpha_{*}(\Gamma / H)_{*}}\right),
    \label{eq:rdapprox}
\end{equation}
where $x_*$ is solved iteratively from:
\begin{equation}
1=\frac{x-3 / 2-\frac{d(\ln g_s)}{d(\ln T)}}{\frac{\Gamma_{e q}}{H}\left[1+\frac{1}{3} \frac{d(\ln g_s)}{d(\ln T)}\right]}  \, , 
\end{equation}
and 
\begin{equation}
\Gamma_{e q} / H=8 \times 10^{34} \left(\frac{m}{\mathrm{GeV}}\right) \left(\frac{ \langle\sigma v\rangle }{\mathrm{cm}^{3} \, \mathrm{s}^{-1}}\right) x^{1 / 2} e^{-x} g_s^{-1 / 2}.    
\end{equation}
We have checked that the approximation is always within 5\% of the value returned by the fully numerical scheme.

\begin{figure}[!ht]
    \centering
    \includegraphics[width=\columnwidth]{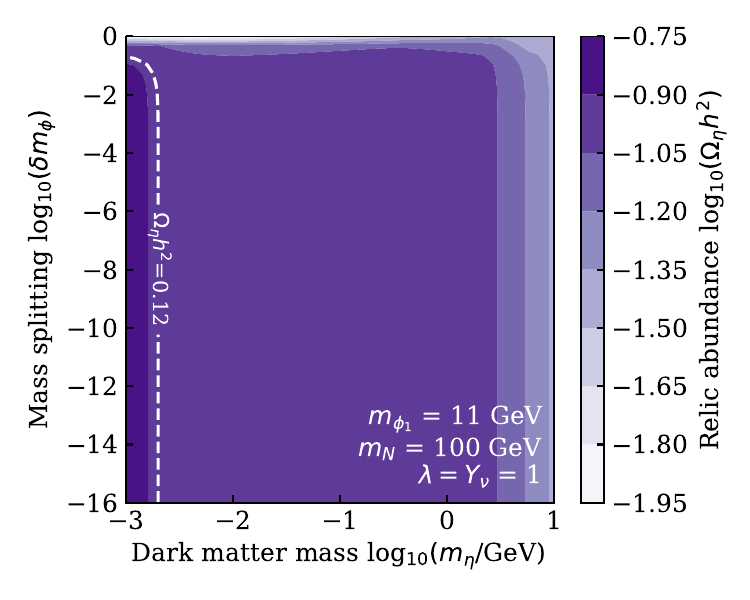}
    \includegraphics[width=\columnwidth]{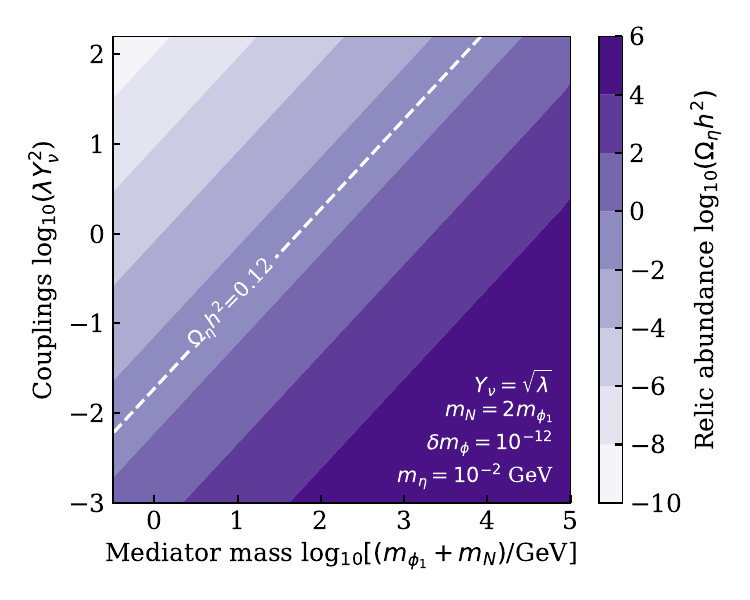}
    \caption{Top: relic abundance as a function of the mass splitting $\delta m\phi$ and dark matter mass $m_\eta$. Fixed $m_{\phi_1}=11$ GeV, $m_N=100$ GeV, $Y_\nu=\lambda=1$. Bottom: relic abundance as a function of the couplings and sum of mediator masses, fixing $Y_\nu = \sqrt{\lambda}, m_N = 2m_{\phi_1}, \delta m_\phi = 10^{-12}$ and $m_\eta = 10$ MeV.}
    \label{fig:mx_lambda_theory}
\end{figure}

We show an example of the relic abundance for a fixed set of parameters in Fig.~\ref{fig:mx_lambda_theory}. In the top panel, we present the mass splitting $\delta m_\phi$ and dark matter mass $m_\eta$. For fixed mediator masses and couplings, the relic abundance is largely independent of the mass splitting for values of less than 0.1 and is inversely proportional to the dark matter mass. In the bottom panel, the relic abundance is a function of the couplings $\lambda Y_\nu^2$ as well as the sum of the mediator masses $m_{\phi_1} + m_N$ for fixed mass splitting and dark matter mass. As expected, dark matter is underproduced at large couplings and low mediator masses, and vice versa.

Instead of fixing parameters as in Fig.~\ref{fig:mx_lambda_theory}, we performed two sets of scans over the parameter space for different relic abundance scenarios. In the \textit{Saturated} scenario, we impose that the model represents the totality of the dark matter density observed today. Second, we allow $\eta$ to be \textit{Underabundant}, i.e. is only a fraction of the full dark matter abundance. 

We use the relic abundance from Planck 2018 \cite{Planck:2018vyg}, $\Omega_{\mathrm{DM}} h^2 = 0.1200 \pm{0.0012}$ at 68\% C.L. and construct a Gaussian likelihood centered at this value. For the Saturated model,
\begin{equation}
-2 \ln \mathcal{L}_{\mathrm{RA}}=\left(\frac{\Omega_{\mathrm{DM}} h^2-\Omega_\eta h^2}{0.0012}\right)^2.
\label{eq:oh2like}
\end{equation}
The Underabundant scenario takes the value of $\Omega_{\mathrm{DM}} h^2$ as an upper limit

\begin{equation}
-2 \ln \mathcal{L}_{\mathrm{RA}}=\left(\frac{\Omega_{\mathrm{DM}} h^2-\Omega_\eta h^2}{0.0012}\right)^2 \Theta(\Omega_\eta h^2-\Omega_{\mathrm{DM}} h^2).
\label{eq:oh2like2}
\end{equation}

\subsection{Late-time annihilation}
\begin{figure}[!ht]
    \centering
    \includegraphics[width=\columnwidth]{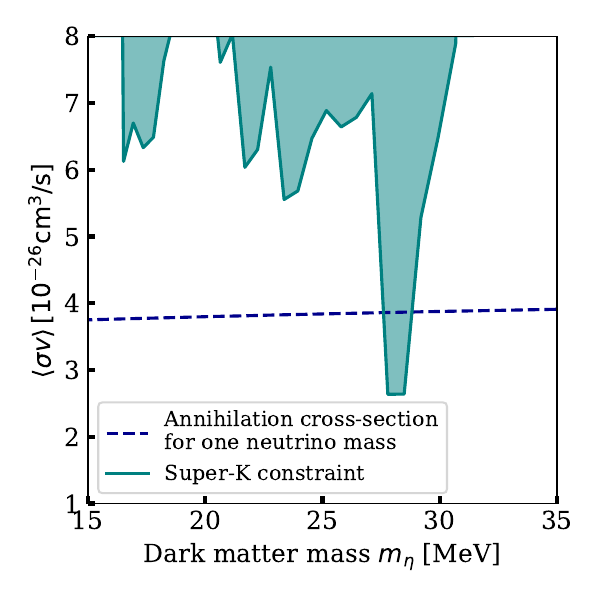}
    \caption{Super-Kamiokande constraint on the s-wave thermally-averaged annihilation cross-section of dark matter to neutrinos.}
    \label{fig:sk_constraint}
\end{figure}

The relic dark matter can self-annihilate at late times, yielding a potential signature in neutrino telescopes, with the majority of the constraining power coming from the direction of the Galactic center, where the flux from dark matter self-annihilation would be highest. Ref. \cite{Arguelles:2019ouk} compiled such constraints over 14 orders of magnitude in dark matter mass. However, only a small window between 25 and 30 MeV is constrained below the relic abundance line (see Fig.~\ref{fig:sk_constraint}). These constraints are based on the limits from a Super-Kamiokande (Super-K) diffuse $\bar \nu_e$ search \cite{Linyan:2018rmj}. 
\begin{equation}
-2 \ln \mathcal{L}_{\mathrm{ann}}=\left(\frac{-\langle\sigma v\rangle}{\frac{1}{1.645} \times \langle\sigma v\rangle_\text{data}}\right)^2\, .
\end{equation}
These are rescaled to account for dark matter annihilations producing a single mass eigenstate ($\nu_1$ or $\nu_3$), and in the underabundant case, are relaxed by a factor of $\Omega_{DM}^2 / \Omega_\eta^2$. 

\subsection{Primordial Nucleosynthesis}
If light dark matter decoupled from the plasma at MeV temperatures, the entropy injection into the neutrino sector will have affected the expansion rate of the Universe, altering the neutron to proton ratio at freeze-out, and thus changing the relic abundances of helium and deuterium. While this is normally parametrized via $N_\text{eff}$, the effective number of neutrinos at BBN, we use the {\fontfamily{lmtt}\selectfont 
AlterBBN} \cite{Arbey_2012, Arbey:2018zfh} software to directly compute the primordial helium and deuterium fractions Y$_{\mathrm{P}}$ and [D/H] in the presence of a dark matter particle with a single degree of freedom, annihilating directly to neutrinos. In our MCMC, we directly implement the $
    -2 \ln \mathcal{L}_{N_\mathrm{eff}} = \chi^2$ values returned by {\fontfamily{lmtt}\selectfont 
AlterBBN}, where the primordial abundances are from Ref.~\cite{PDG2018}:
\begin{equation}
\begin{aligned}
Y_{\mathrm{P}} & =0.2450 \pm 0.003, \\
{[\mathrm{D}] /[\mathrm{H}] } & =(2.569 \pm 0.027) \times 10^{-5}.
\end{aligned}
\end{equation}

As mentioned in the introduction, scotogenic models have been linked to leptogenesis. Our effective low-energy lagrangian Eq.~\eqref{eq:effective_lagrangian} does not directly couple to the charged leptons, so a full baryogenesis and leptogenesis analysis is not possible. Models of leptogenesis can produce a large asymmetry in the neutrino sector \cite{Simha:2008mt}.  We do not include such a constraint, since, again, it would require a more UV-complete model, and including all three neutrino flavors simultaneously.

\subsection{Dark matter-neutrino elastic scattering}

Elastic scattering between relic neutrinos and dark matter can lead to a suppression of the matter power spectrum at large wavenumber $k$ and erase small-scale structure through collisional damping \cite{Boehm:2000gq,Boehm:2001hm,Boehm:2004th, Wilkinson:2014ksa,Escudero:2015yka,Crumrine:2024sdn,Heston_2024}. The elastic scattering cross section can be obtained from the self-annihilation rate via crossing symmetry. This computation is detailed in App. \ref{app:scattering}. 

\begin{figure}[!ht]
    \centering
    \includegraphics[width=\columnwidth]{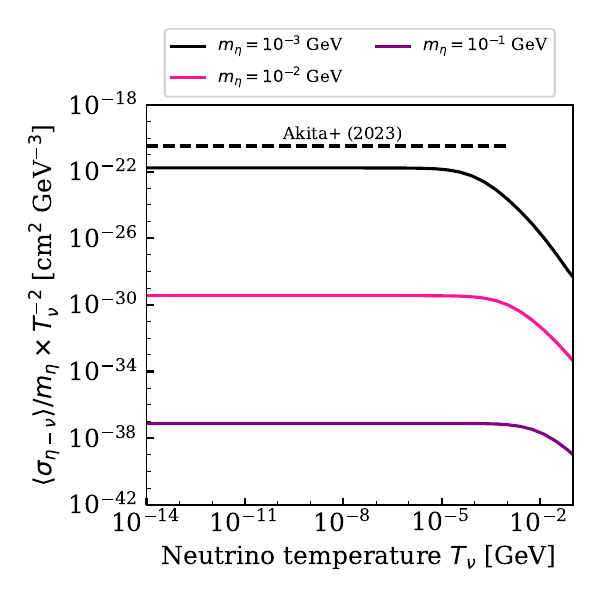}
    \caption{Scattering opacity divided by the neutrino temperature squared as a function of neutrino temperature for different dark matter masses. Fixed $m_{\phi_1}=m_{\phi_2}=10m_\eta$, $m_N=100 m_\eta$, and $\lambda=Y_\nu=0.5$. At the temperature of the C$\nu$B ($T_\nu \approx 1.7 \times 10^{-13}$  GeV), the thermally-averaged scattering cross-section is fully $d$-wave. We implement the Akita+ \cite{Akita:2023yga} upper limit since is always the dominant one for this model, though these scattering limits will turn out to be subdominant to other constraints discussed in the text.}
    \label{fig:rescaled_opacity}
\end{figure}

At low temperatures, the leading term for the thermally-averaged dark matter-neutrino scattering cross section is  proportional to $T^2$. We use the constraint on the opacity parameter $\langle\sigma_{\eta-\nu}\rangle/m_\eta$ from Ref. \cite{Akita:2023yga}, which is the strongest limit in the literature applicable to this model. We take a half-Gaussian likelihood  centered at zero with a 2-sigma width of $\langle\sigma_{\eta-\nu}\rangle/m_\eta\leq 10^{-46} \, \text{cm}^2/\text{GeV}$. By convention, this is defined as the cross section at today's relic neutrino temperature $T_0 \simeq 1.7\times 10^{-4}$ eV. Then,

\begin{equation}
 -2 \ln \mathcal{L}_{\mathrm{sc}}=\left(\frac{-\langle\sigma_{\eta-\nu}\rangle/m_\eta}{0.5 \times 10^{-46} \, \text{cm}^2\text{/GeV}}\right)^2 \, .
\end{equation}
Many recent references have revisited these constraints, but mainly focus on cross sections that are constant with $T$ \cite{Ferrer:2022kei,Cline:2022qld,Fujiwara:2023lsv}. As we will find in Sec. \ref{sec:results}, these elastic scattering constraints will always be subdominant to the others discussed here, so they will not be explored further. 

\subsection{Neutrino masses}
We wish to test the viable parameter space in this model, while retaining the simplicity of a one-flavor model. We thus consider two scenarios, looking at the lightest and heaviest neutrino masses separately. We comment on how these may be combined in Sec.~\ref{sec:discussion}.

Ref.~\cite{GAMBITCosmologyWorkgroup:2020rmf} performed a global fit over cosmological parameters, with data from \textit{Planck} \cite{Planck:2018vyg}, Baryon Acoustic Oscillation (BAO) \cite{2011MNRAS.416.3017B,BOSS:2016wmc,Ata:2017dya,Bautista:2017wwp,Blomqvist:2019rah,DES:2017rfo}, and oscillation parameters from NuFit 5.1 \cite{Esteban:2018azc}. They found an upper limit for the mass of the lightest neutrino  of 0.037 eV at 95\% C.L., assuming Normal Ordering (NO). For the \textit{light neutrino} scenario we implement the likelihood:
\begin{equation}
-2 \ln \mathcal{L}_{m_{\nu_1}}=\left(\frac{-m_{\nu_0}}{0.0185\, \text{eV}}\right)^2 \,.
\label{eq:lnlike_light_nu}
\end{equation}
This is unbounded from below, and the possibility of a near-zero mass will end up producing the majority of the weight in the parameter space.

For the \textit{heavy neutrino} scenario, we vary the mass of the lightest neutrino $m_{\nu_1}$ using Eq.~\eqref{eq:lnlike_light_nu} as the prior, we calculate $m_{\nu_3}$ with Eq.~\eqref{eq:neutrino_mass} and obtain $\Delta m_{31}^2 = m_{\nu_3}^2 - m_{\nu_1}^2$ whose $\chi^2_{\Delta m_{31}^2}$ is available from oscillation data with {\fontfamily{lmtt}\selectfont 
NuFit} \cite{Esteban:2020cvm}:

\begin{equation}
-2 \ln \mathcal{L}_{\Delta m_{31}^2} = \chi^2_{\Delta m_{31}^2} \,.
\label{eq:lnlike_heavy_nu}
\end{equation}

\subsection{Z boson and kaon decays}

As pointed out in \cite{Alvey:2019jzx}, the coupling in Eq.~\eqref{eq:effective_lagrangian} contributes to new decay modes for mesons. For the lightest neutrino case, we choose the bounds on the coupling $Y_\nu$ to electron and muon neutrinos and we transform them from the flavor to the mass basis.

In the case that $m_N+m_{\phi_1} \leq m_{K} \simeq \mathcal{O}(500 \,\mathrm{MeV})$ we adopt the constraint on the kaon decay channels $K^{+} \rightarrow \ell^{+} +N_R+\phi_1$ obtained by Refs.~\cite{Lessa_2007,Farzan:2010wh} with data from the SPEC, OSPK and KLOE experiments and improved by the inclusion of NA48/2 data on the branching ratio of kaon decays to electrons and muons:
\begin{equation}
    \left|Y_{\nu e}\right| \lesssim 3 \times 10^{-3} \, ,
    \label{eq:kaon_decay_e}
\end{equation}
\begin{equation}
    \left|Y_{\nu \mu}\right| \lesssim 10^{-2} \, ,
    \label{eq:kaon_decay_mu}
\end{equation}
at 90\% C.L. in the flavor basis. We transform this to the mass basis via $\left|Y_{\nu l}\right| = \left|Y_{\nu}\right| |U_{li}|$ for $i = 1$ (light neutrino) or $i = 3$ (heavy neutrino). 
where $|U_{li}|$ are the elements of the PMNS matrix and we choose the best-fit values from {\fontfamily{lmtt}\selectfont 
NuFit}.

These are more conservative bounds compared to Ref. \cite{Pasquini:2015fjv} for example, but we will see that they are more than enough to close the parameter space in combination with the relic abundance. The likelihood for the kaon decay is
\begin{equation}
-2 \ln \mathcal{L}_{K}=\left(\frac{-Y_{\nu e}}{\frac{1}{1.645} \times 3 \times 10^{-3}}\right)^2 + \left(\frac{-Y_{\nu \mu}}{\frac{1}{1.645} \times 10^{-2}}\right)^2 \, .
\label{eq:lnlike_k_decay}
\end{equation}
We also considered $D$ meson decay constraints, which are present when $m_K<m_{\phi_1}+m_{N}<m_D$, where $m_D$ is the mass of the $D$ meson. We finds that these are subdominant to Z decay constraints (which we discuss next) and we therefore do not include them in this analysis.

Lastly, we derive constraints from the Z boson decay. The expression for the decay width is unwieldy, and can be found in Appendix \ref{app:zdecay}. We take a conservative bound assuming the new decay channel does not contribute beyond the total uncertainty on the Z boson decay width, 0.0023 GeV at 95\% C.L. \cite{ParticleDataGroup:2024cfk}. The likelihood is thus
\begin{equation}
-2 \ln \mathcal{L}_{Z}=\left(\frac{-\Gamma_Z}{0.5 \times 0.0023 \, \mathrm{GeV}}\right)^2
\end{equation}

then the total decay likelihood is
\begin{equation}
    -2 \ln \mathcal{L}_\text{decays} = -2 \ln \mathcal{L}_{Z} -2 \ln \mathcal{L}_{K}.
\end{equation}

\begin{figure}[!ht]
    \centering
    \includegraphics[width=\columnwidth]{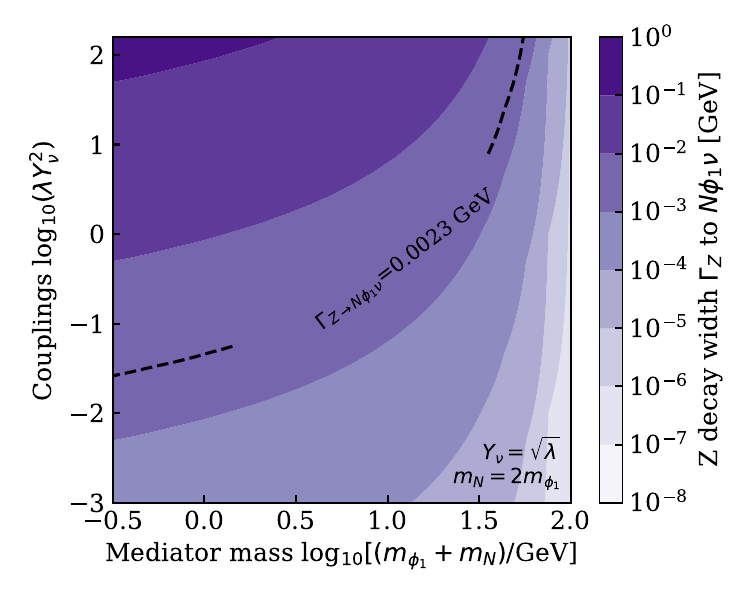}
    \caption{Z boson width constraint on the new decay channel $N\phi_1\nu$ assuming it does not contribute more than the total uncertainty of 0.0023 GeV at 95\% C.L. of the Z boson decay width depicted by the black dashed line. Fixed relationship between the couplings $Y_\nu = \sqrt{\lambda}$ and the mediator masses $m_N = 2m_{\phi_1}$.}
    \label{fig:z_decay_constraint}
\end{figure}

\section{Results: Global analysis} \label{sec:results}

We perform a set of MCMC parameter scans using the \texttt{emcee} package. We present results both in terms of Bayesian marginalised posteriors (showing 95.45\% credible regions, C.R.), and as frequentist profile likelihood ratios (showing 95.45\% confidence limits, C.L.). In each set of figures that we show, the results of the parameter space analysis will be shown as contours, with coloured regions indicating the most limiting constraints coming from each side. We present plots in terms of four cases: saturated dark matter and the mass of the lightest neutrino (Saturated DM + Light $\nu$), saturated dark matter and the mass of the heaviest neutrino (Saturated DM + Heavy $\nu$), underabundant dark matter and the mass of the lightest neutrino (Underabundant DM + Light $\nu$), and underabundant dark matter and the mass of the heaviest neutrino (Underabundant DM + Heavy $\nu$). 
These constraints will display some small-scale ``noise'': this is due to a combination of sampling resolution, and artefacts of binning and interpolation, and is not physically meaningful. We will find that the mass hierarchy that we impose to stabilise the dark matter, along with the requirement of perturbativity on the couplings in combination with all of the constraints mean that the parameter space is bounded from all sides, with the exception of the mass splitting (in the case of the lightest neutrino) since it is allowed to be massless, and thus we are simply left with $\delta m_\phi \ge 0$. \\

\begin{figure}[!ht]
    \centering
    \includegraphics[width=\columnwidth]{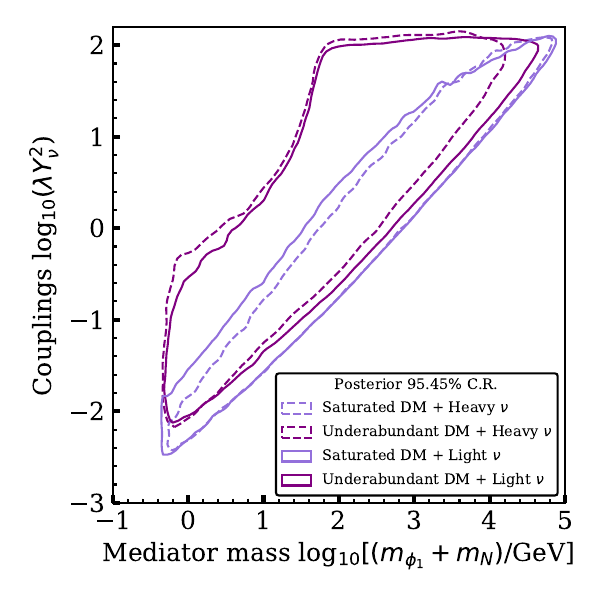}
      \includegraphics[width=\columnwidth]{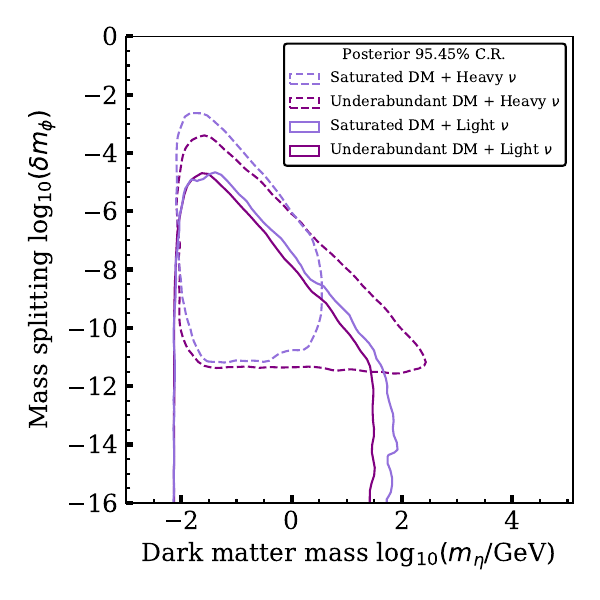}
    \caption{95.45\% C.R. posterior constraints from the four MCMC parameter scans. Top: couplings-mediator mass plane; bottom: mass splitting--dark matter mass. Line styles represent coupling to the heaviest neutrino eigenstates ($\nu_3$, dashed) and lightest ($\nu_1$, solid), assuming normal ordering. The light color shows dark matter $\eta$ that saturates the DM relic abundance, while the dark color is allowed to be underabundant.}
    \label{fig:couplings_mediators_all}
\end{figure}

Fig.~\ref{fig:couplings_mediators_all} summarizes our results, with the 95\% allowed regions from the four different scans, presented as the combination of the couplings $\lambda Y_\nu^2$ against the sum of mediator masses $m_{\phi_1}+m_N$ (top panel) and the mass splitting $\delta m_\phi$ versus the dark matter mass $m_\eta$ (bottom panel). These four scenarios clearly share some features: low dark matter masses are excluded by BBN, which in turn restricts the mediator masses via the dark matter stability condition. The mediator masses cannot be arbitrarily large, since this should be compensated by correspondingly large couplings, limited by perturbativity. The most striking difference in the lower panel comes from the choice of heavy or light neutrino; in the latter case, the mass splitting can be vanishingly small, as the lightest neutrino may still be massless. 

\begin{figure*}[!ht]
    \centering
        \includegraphics[width=\columnwidth]{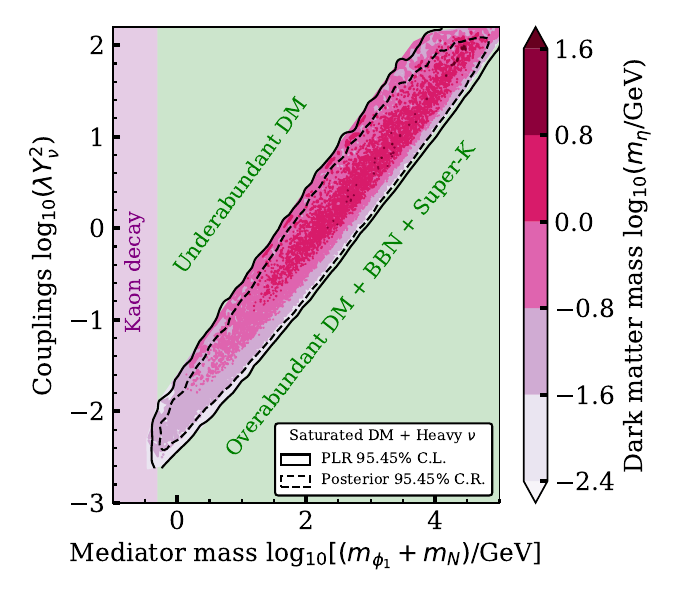}
        \includegraphics[width=\columnwidth]{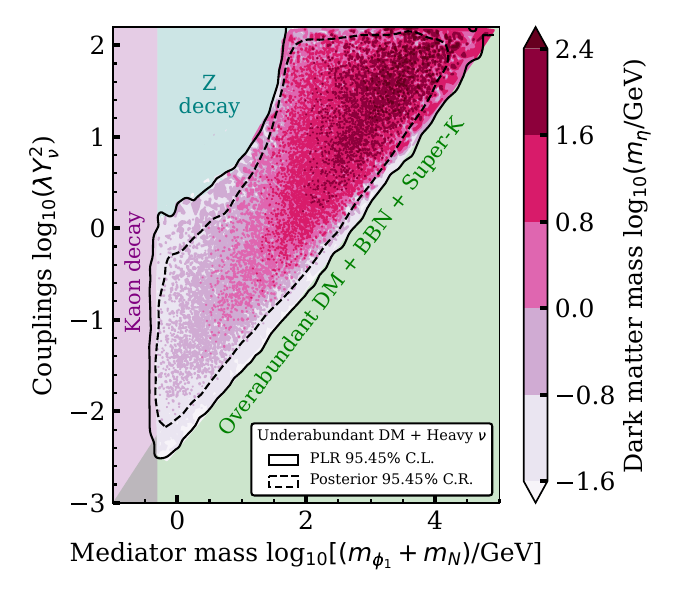}
        \includegraphics[width=\columnwidth]{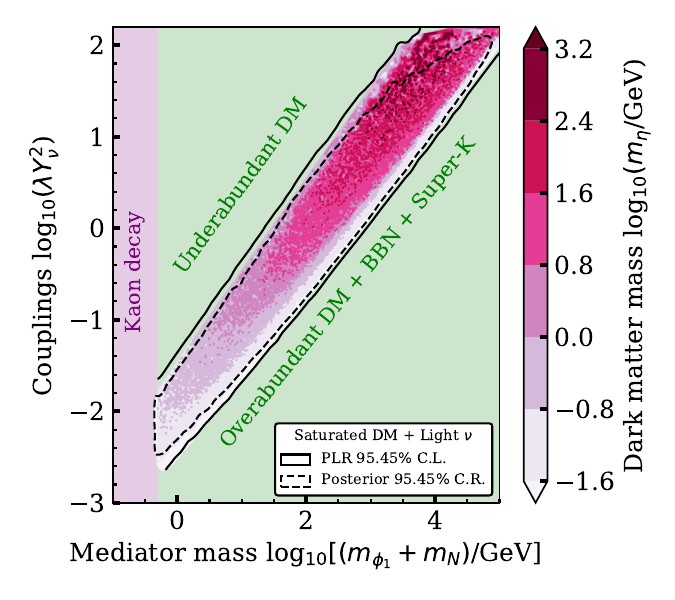}
        \includegraphics[width=\columnwidth]{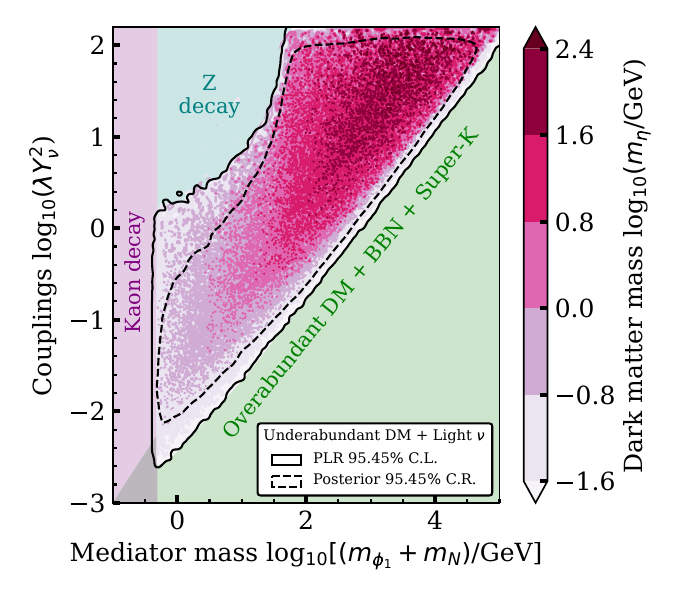}

    \caption{Parameter space of the coupling combination $\lambda Y_\nu^2$ and the sum of the mediator masses $m_{\phi_1} + m_N$ as a function of the dark matter mass $m_\eta$. The solid lines represent the 95.45\% C.L. of the PLR and the dashed lines 95.45\% C.R. of the posteriors for the cases when the dark matter candidate represents the total observed relic abundance (left panels) and when is underabundant (right panels). As well as the cases where the mass of the heaviest (top panels) and lightest (bottom panels) neutrino is reproduced. The shaded regions illustrate the dominant constraint on each side, see text for details.}
    \label{fig:ynu_mediators_2sigma}
\end{figure*}

\begin{figure*}[!ht]
    \centering
    \includegraphics[width=\columnwidth]{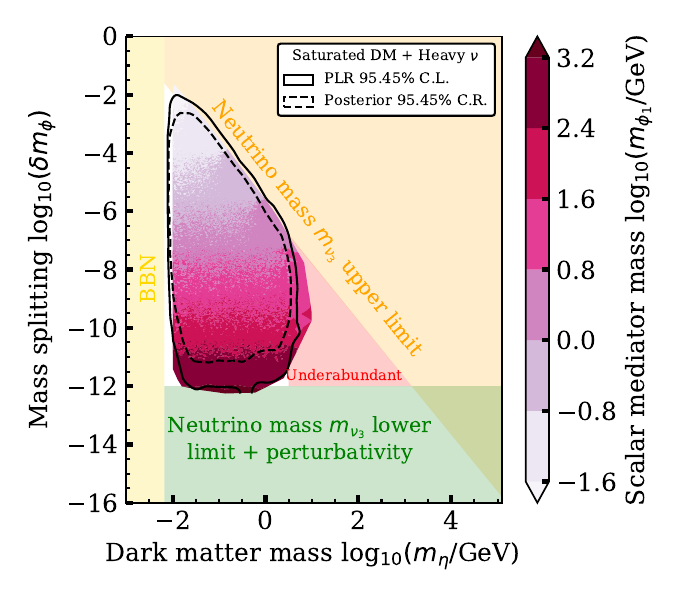}
    \includegraphics[width=\columnwidth]{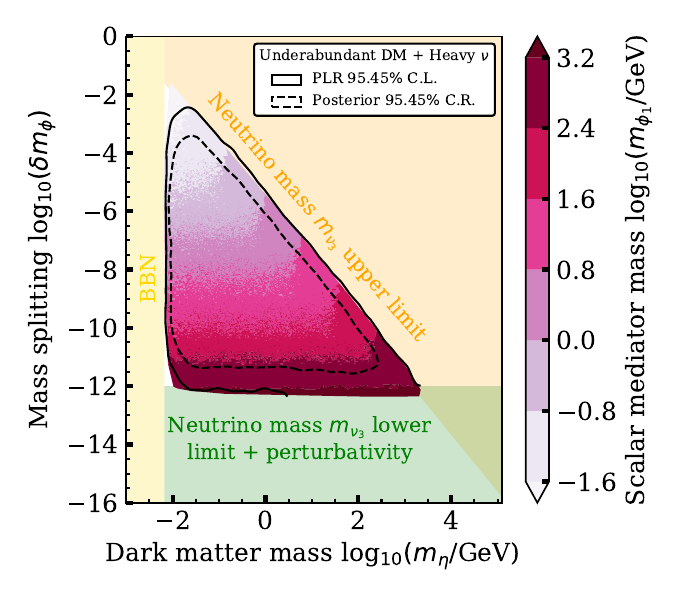}
    \includegraphics[width=\columnwidth]{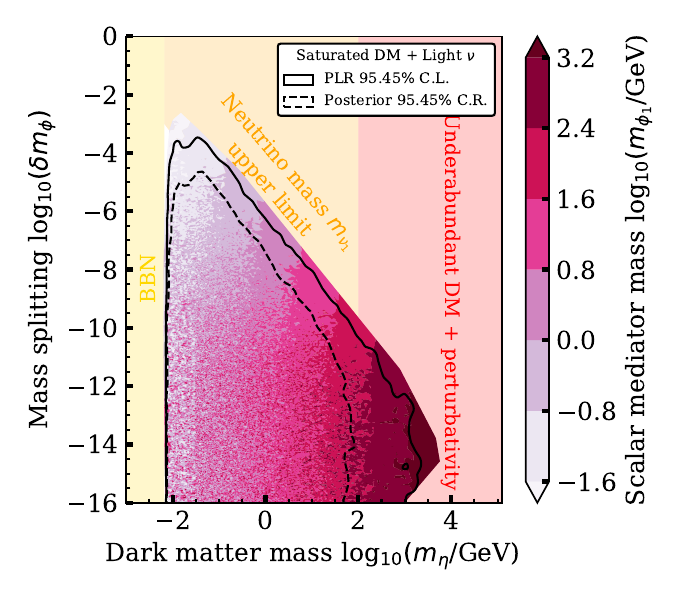}
    \includegraphics[width=\columnwidth]{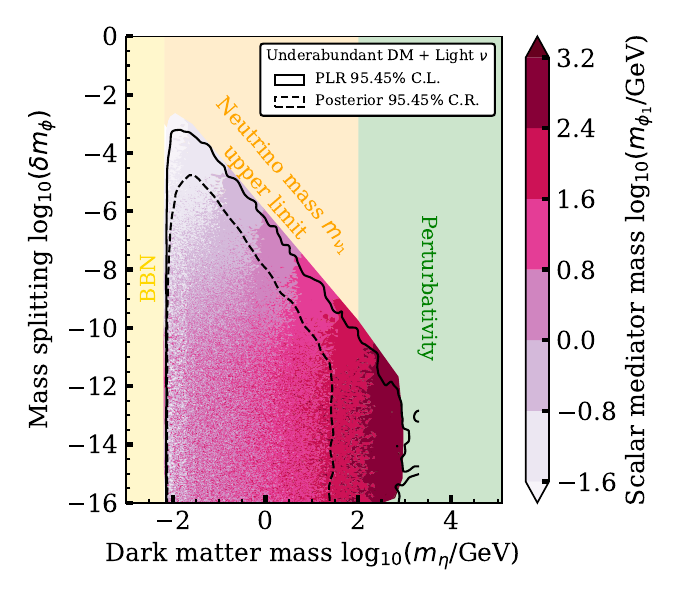}

    \caption{Parameter space of the mass splitting $\delta m_\phi$ and the dark matter mass $m_\eta$ as functions of the scalar mediator mass $m_{\phi_1}$. The solid lines represent the 95.45\% C.L. of the PLR and the dashed lines 95.45\% C.R. of the posteriors for the cases when the dark matter candidate represents the total observed relic abundance (left panels) and when is underabundant (right panels). As well as the cases where the mass of the heaviest (top panels) and lightest (bottom panels) neutrino is reproduced.}
    \label{fig:deltamphi_meta_2sigma}
\end{figure*}

These constraints are shown more explicitly in Figs. \ref{fig:ynu_mediators_2sigma}, corresponding to the $\lambda Y^2_\nu - m_{\phi_1}+m_N$ plane, and \ref{fig:deltamphi_meta_2sigma}, showing the $\delta m_\phi - m_\eta$ plane. In these figures, we show the 95.45\% confidence level (frequentist) profile likelihood ratio contour as a solid line, and the 95.45\% (bayesian) credible region as a dashed line. The two generally trace each other well. \\

Fig.~\ref{fig:ynu_mediators_2sigma}  shows the plane of the couplings versus mediator masses for the four different cases, and the color scale indicates the dark matter mass $m_\eta$ for each of the samples. The parameter space is clearly bounded from all sides by the constraints, as described above, represented by shaded regions. These are overlaid to indicate the dominant constraint in each region. 
In all cases, the couplings and mediator masses are bounded from below by a combination of the relic abundance, BBN, annihilation cross-section and kaon decay. As the mediator masses and couplings increase, larger mass splittings and lower dark matter masses are needed to keep dark matter from being overproduced. The low DM masses in turn are constrained in a small range by the Super-K limit on the annihilation cross-section and below by BBN. 

We choose to focus on a WIMP-like model where dark matter is produced thermally through a freeze-out mechanism. Therefore, it is natural for bigger couplings to be preferred, but the parameter space is closed at $m_{\phi_1} + m_N < 500$ MeV by the constraint from kaon decay. As mentioned in the introduction, we opt for concentrating on this region and leave the question of other dark matter production mechanisms like freeze-in for future work.

The Saturated DM + Heavy $\nu$ (top left) and Saturated DM + Light $\nu$ (bottom left) cases are bounded from above by the perturbativity of the couplings in combination with the DM relic abundance. The bounds are slightly different for both cases due to limits on the mass of the heaviest and lightest neutrino, respectively. The Underabundant DM + Heavy $\nu$ (top right) and Underabundant DM + Light $\nu$ (bottom right) cases are bounded from above by the perturbativity of the couplings in combination with the limits on the Z decay to $N \phi_1 \nu$.

As mentioned earlier, elastic DM-$\nu$ scattering constraints from cosmology are always subdominant, and do not play an important role for this model. 

In Fig.~\ref{fig:deltamphi_meta_2sigma} all panels show that the model naturally favors lower ($\sim 10$ MeV-GeV) mass dark matter in order to generate non-zero neutrino masses and/or be consistent with relic abundance and perturbativity constraints. Furthermore, the mass splitting is bounded from above by the combination of BBN and the upper limit of the neutrino masses. In the Saturated DM + Heavy $\nu$ (top left) and Underabundant DM + Heavy $\nu$ (top right) cases, the mass splitting is bounded from below by the lower limit on the mass of the heaviest neutrino and the perturbativity of the couplings. 

In these cases, we notice that the mass splitting is roughly inversely proportional to the scalar mediator mass (as seen in the pink color scale), so smaller mass splittings lead to larger mediator masses in order to reproduce the mass of the heaviest neutrino. Additionally, larger couplings are needed, which leads to the perturbativity limit. In the Saturated DM case, the upper limit on the DM mass is tighter since it is not allowed to be underproduced. 

In the Saturated DM + Light $\nu$ (bottom left) and Underabundant DM + Light $\nu$ (bottom right) cases the mass splitting is not bounded from below since the neutrino is allowed to be massless as previously mentioned. For the same reason, we note that the relationship of the scalar mediator mass (color bar) is not set by the mass splitting as in the previous cases, but is proportional to the DM mass through the hierarchy $m_\eta<m_{\phi_1}$. Therefore, the upper limit on the DM mass in these cases is set by the perturbativity of the couplings and, in the saturated case, in combination with the DM not being underabundant.
\section{Discussion and Conclusion} \label{sec:discussion}

We have performed a detailed phenomenological analysis of a model in which scalar dark matter interacts with neutrinos, primarily at next-to-leading order in perturbation theory. 
This scenario self-consistently provides an explanation for light thermal relic dark matter as well as a neutrino mass generation mechanism. We have shown that various cosmological, astrophysical, and experimental observables bound this model, on all sides, leaving only a finite volume of parameter space available. 
We find that the dark matter relic abundance and perturbativity constraints naturally lead to a sub-GeV dark matter candidate, constraining the model from the right. At the same time, BBN prevents the dark matter from being lighter than $\sim 10$ MeV, while heavy meson decays require the sum of the mediator masses to be heavier than $\sim 500$ MeV, bounding the parameter space from the left.
Finally, an upper limit on the neutrino mass and rare Z-boson decays bound the space from above.

The bounds on this dark sector coupling to the lightest and heaviest neutrinos are fairly robust, and the constraints on the parameter space are only relaxed by a small amount when $\eta$ is allowed to be underabundant. Taken together, these observations hint that a full 3-flavor model should not be so different. Such an analysis would come at the cost of many more free parameters and correlations, but the generalization is in principle straightforward. Additional constraints (on the 3-flavor model)---and opportunities---would of course arise from neutrino oscillation, these are beyond the scope of this work. Similarly, a more UV-complete model could potentially tie in with leptogenesis, unifying three major questions in particle cosmology.

A remaining question, then, is how detectable such an extension of the standard model is in the near future. A more precise measurement of the relic abundance $\Omega_{DM} h^2$ will of course narrow the range of couplings and mediator masses, while next-generation oscillation parameter measurements will help pin down the mass splittings. Upcoming experiments, such as DUNE \cite{Buckley:2022btu}, JUNO \cite{Arguelles:2019ouk}  and Hyper-Kamiokande \cite{Bell:2020rkw}, could offer improvements by providing sensitivity to dark matter-neutrino annihilation cross-sections; a neutrino line signal would indeed be a smoking gun for a neutrinophillic dark sector.

Precision experiments that search for rare kaon decays, such as NA62 \cite{NA62:2020fhy} and KOTO \cite{KOTO:2020prk}, could close in on the parameter space at sub-GeV masses. This would directly impact the viability of the underabundant scenario, but the saturated cases would remain unaffected. The same would likely be true of precision $Z$ decay measurements at a future $e^+e^-$ (or $\mu^+\mu^-$) collider \cite{FCC:2018evy,MuonCollider:2022xlm}. Additionally, lepton number violating interactions in the neutrino sector could have sizable effects in core-collapse supernovae explosions and be testable in future detections \cite{Suliga:2024oby}.

Our main conclusions are fairly generalizable, making our effective model, and indeed any UV completion that reduces to a similar form, a prime target for future searches.


\section*{Acknowledgements}

We thank Richard Ruiz, Cristina Mondino, Matheus Hostert, Yasaman Farzan, Vedran Brdar, Martín de los Rios and the TheDeAs group for useful discussions. All authors acknowledge support from the Arthur B. McDonald Canadian Astroparticle Physics Research Institute. KMC is supported by ``la Caixa" Foundation through the fellowship LCF/BQ/DI22/11940024 and by the Spanish Agencia Estatal de Investigación through the grants PID2021-125331NB-I00 and CEX2020-001007-S. During completion of this work, GM was supported in part by the UC office of the President through the UCI Chancellor's Advanced Postdoctoral Fellowship, the U.S. National Science Foundation under Grant PHY-2210283 and the Natural Sciences and Engineering Research Council of Canada (NSERC).
ACV is supported by NSERC and the province of Ontario through an ERA, with equipment funded by the Canada Foundation for Innovation and the Province of Ontario, and housed at the Queen’s Centre for Advanced Computing.
Research at Perimeter Institute is supported by the Government of Canada through the Department of Innovation,Science, and Economic Development, and by the Province of Ontario. 

\bibliography{loops}

\appendix

\onecolumngrid
\section{Amplitudes and cross-sections}
\label{app:crosssections}

We calculate the amplitudes both in terms of $\Phi$ and in terms of $\phi_1$ and $\phi_2$. The annihilation and scattering amplitudes themselves are not greatly changed by the introduction of very small neutrino masses, so we neglect their contribution but keep the mass splitting between the real scalar and the pseudo-scalar. 

The amplitude is:

\begin{equation}
    \mathcal{M}=-\frac{1}{16\pi^2}\lambda Y_\nu^2 m_N C_0(0,0,s,m_\Phi^2,m_N^2,m_\Phi^2)\Bar{u}(k_2)P_{R,L}v(k_1)
\end{equation}
where $s$ is the Mandelstam variable related to the centre of mass energy as the squared of the sum of the incoming momenta, $C_0(p_1^2, p_2^2,\left(p_1+p_2\right)^2, m_1^2, m_2^2, m_3^2)$ is the scalar triangle integral as defined by the {\fontfamily{lmtt}\selectfont  LoopTools} \cite{looptoolsHahn_1999} convention, also know as the scalar three-point function. Using {\fontfamily{lmtt}\selectfont  Package-X} \cite{Patel_2015} we find an analytical expression for the scalar integral $C_0$:
\begin{equation}
\begin{split}
&  C_0\left(0,0,s,m_\Phi^2,m_N^2,m_\Phi^2\right) = \\
   & \frac{\text{DiLog}\left(\frac{2 \left(\text{$m_N$}^2-\text{$m_\phi$}^2\right)}{2 \text{$m_N$}^2-\sqrt{s \left(s-4 \text{$m_\phi$}^2\right)}-2 \text{$m_\phi$}^2+s},s\right)}{s}+\frac{\text{DiLog}\left(\frac{2 \left(\text{$m_N$}^2-\text{$m_\phi$}^2\right)}{2 \text{$m_N$}^2+\sqrt{s \left(s-4 \text{$m_\phi$}^2\right)}-2 \text{$m_\phi$}^2+s},s\right)}{s} \\
   & -\frac{\text{DiLog}\left(\frac{2 \left(\text{$m_N$}^2-\text{$m_\phi$}^2+s\right)}{2 \text{$m_N$}^2-\sqrt{s \left(s-4 \text{$m_\phi$}^2\right)}-2 \text{$m_\phi$}^2+s},s\right)}{s}-\frac{\text{Li}_2\left(\frac{2 \left(\text{$m_N$}^2-\text{$m_\phi$}^2+s\right)}{2 \text{$m_N$}^2-2 \text{$m_\phi$}^2+s+\sqrt{s \left(s-4 \text{$m_\phi$}^2\right)}}\right)}{s} \\
   & +\frac{\text{Li}_2\left(\frac{\left(\text{$m_N$}^2-\text{$m_\phi$}^2\right) \left(\text{$m_N$}^2-\text{$m_\phi$}^2+s\right)}{\text{$m_N$}^4-2 \text{$m_\phi$}^2 \text{$m_N$}^2+s \text{$m_N$}^2+\text{$m_\phi$}^4}\right)}{s}-\frac{\text{Li}_2\left(\frac{\left(\text{$m_N$}^2-\text{$m_\phi$}^2\right)^2}{\text{$m_N$}^4-2 \text{$m_\phi$}^2 \text{$m_N$}^2+s \text{$m_N$}^2+\text{$m_\phi$}^4}\right)}{s}
\end{split}
\end{equation}

where $\text{DiLog}(z,s) = \operatorname{Li_2}(z)$ if $z$ is a complex number or a real number less than one. For the real case $z>1$, the sign of $s$ determines on which side of the branch to evaluate:

\begin{equation}
\text{DiLog}(z,s) =
\begin{cases}
    \operatorname{Li_2}(z) & \text{if } s < 0 \\
    -\frac{1}{2}\log\left(\frac{1}{1-z}\right)^2 - \operatorname{Li_2}\left(\frac{z}{-1+z}\right) & \text{otherwise}
\end{cases}
\end{equation}

In the annihilation case, the two incoming momenta are the dark matter particles. 

The full spin-averaged squared amplitude of real scalar dark matter annihilating into neutrinos is:

\begin{equation}
    |\overline{\mathcal{M}}|^2_{\text{ann}}=\frac{1}{128\pi^4}\lambda^2 Y_\nu^4 m_N^2 s |C_0(0,s,0,m_N^2,m_\Phi^2,m_\Phi^2)|^2
    \label{eq:amplitude1}
\end{equation}

For this work, we use the following parametrization of the amplitude in terms of $\phi_1$ and $\phi_2$:

\begin{equation}
    |\overline{\mathcal{M}}|^2_{\text{ann}} = \frac{\lambda^2 m_N^2 s Y_\nu^4}{512 \pi ^4} \left(C_0(\phi_1)+C_0(\phi_2)\right)\left(C_0^*(\phi_1)+C_0^*(\phi_2)\right)
    \label{eq:amplitude2}
\end{equation}

where $C_0(\phi_i)=C_0\left(0,0,s,m_{\phi_i}^2,m_N^2,m_{\phi_i}^2\right)$. In the case where $m_{\phi_1}=m_{\phi_2}=m_\Phi$, Eq. \eqref{eq:amplitude2} reduces to Eq. \eqref{eq:amplitude1}. Therefore, when the mass splitting between $\phi_1$ and $\phi_2$ is very small, the phenomenology of the annihilation process is given by Eq. \eqref{eq:amplitude1}.

We perform an approximation to the thermally averaged annihilation cross-section using a low velocity expansion as in \cite{Wells1994}:

\begin{equation}
\sigma v_{\mathrm{rel}}=a_{\mathrm{rel}}+b_{\mathrm{rel}} v_{\mathrm{rel}}^2+c_{\mathrm{rel}} v_{\mathrm{rel}}^4+\cdots
\end{equation}

We take the s-wave approximation $\sigma v_{\mathrm{rel}}=a_{\mathrm{rel}}$ and find that:


\begin{align}
    \langle\sigma v\rangle &=\frac{1}{32\pi m_\eta^2}|\overline{\mathcal{M}}|^2 \\
&= \frac{1}{32\pi m_\eta^2} \frac{\lambda^2 m_N^2 4m_\eta^2 Y_\nu^4 |C_0\left(0,0,4m_\eta^2,m_\Phi^2,m_N^2,m_\Phi^2\right)|^2}{128 \pi ^4} \\
 & = \frac{\lambda^2 Y_\nu^4 m_N^2}{1024\pi^5} |C_0\left(0,0,4m_\eta^2,m_\Phi^2,m_N^2,m_\Phi^2\right)|^2
\label{eq:ann_sigmav_swave}
\end{align}

or in terms of $\phi_1$ and $\phi_2$:

\begin{equation}
    \langle \sigma v \rangle = \frac{\lambda^2 Y_\nu^4 m_N^2}{4096\pi^5} \left(C_0(\phi_1)+C_0(\phi_2)\right)\left(C_0^*(\phi_1)+C_0^*(\phi_2)\right)
\label{eq:ann_sigmav_swave_2}
\end{equation}
To validate the accuracy of our approximation, we compared it to the full thermal average.

\subsection{Dark matter-neutrino elastic scattering}
\label{app:scattering}
To obtain the elastic scattering cross-section we use a crossing symmetry from the Mandelstam variable $s \longrightarrow t$ and multiply by 1/2 to account for the spin-average factor of the incoming neutrino. The relationship between the squared amplitudes is 

\begin{equation}
|\overline{\mathcal{M}}|^2_{\text{sc}} = -\frac{1}{2}|\overline{\mathcal{M}}|^2_{\text{ann}}(s\longrightarrow t).
\end{equation}

To calculate the elastic scattering cross-section we Taylor-expand the amplitude around small momentum transfer by setting the Mandelstam variable $t$ as $t=0$, and truncate to order $t^2$ since this replicates the behavior of the cross-section at low neutrino energies (Eq. \ref{eq:C0expansion}). 

\begin{equation}
\begin{split}
    &C_0\left(0,0,t,m_\Phi^2,m_N^2,m_\Phi^2\right) \approx a+bt+ct^2 \\
    &\Delta = m_N^2-m_\Phi^2 \\
    & a = 1/\Delta-(m_N^2 \log(m_N^2/m_\Phi^2)/\Delta^2) \\
    & b = -(2 m_N^4+5 m_N^2 m_\Phi^2-m_\Phi^4)/(12 m_\Phi^2 \Delta^3)+m_N^4 \log(m_N^2/m_\Phi^2)/(2\Delta^4) \\
    & c = -(3 m_N^8-27 m_N^6 m_\Phi^2-47 m_N^4 m_\Phi^4+13 m_N^2 m_\Phi^6-2 m_\Phi^8)/(180 m_\Phi^4 \Delta^5) \\
    & -m_N^6 \log(m_N^2/m_\Phi^2)/(3 \Delta^6)
\end{split}
\label{eq:C0expansion}
\end{equation}

This is a good approximation because we are using constraints where the neutrino energies of interest are those of the C$\nu$B, at around $1.67\times10^{-4}$ eV.

In the frame of the dark matter, that is basically stationary, the differential cross-section is \cite{Arguelles_2017}:

\begin{equation}
    \frac{d \sigma}{d \cos \theta}=\frac{1}{2 E_\nu} \frac{1}{2 m_\chi} \frac{1}{8 \pi} \frac{E_\nu^{\prime 2}}{E_\nu m_\chi}|\overline{\mathcal{M}}|^2_\text{sc}
\label{eq:dcs_scattering}
\end{equation}

where $E_\nu^{\prime}$ is the outgoing neutrino energy, $E_\nu$ is the incoming energy and $x \equiv \cos \theta$ is the scattering angle, related to each other by:

\begin{equation}
E_\nu^{\prime}=\left(\frac{1}{E_\nu}+\frac{1}{m_\chi}(1-x)\right)^{-1},
\end{equation}

In this frame, the Mandelstam variables are:

\begin{equation}
\begin{aligned}
s &=2 E_\nu m_\chi+m_\chi^2, \\
t &=-2 E_\nu E_\nu^{\prime}(1-x), \\
u &=-2 m_\chi E_\nu^{\prime}+m_\chi^2 .
\end{aligned}
\end{equation}

We obtain the elastic scattering cross-section by integrating Eq.(\ref{eq:dcs_scattering}) over $\cos{\theta}$.

Using Eq.(\ref{eq:amplitude1}) and by remembering that $|\overline{\mathcal{M}}|^2_{\text{sc}} = -\frac{1}{2}|\overline{\mathcal{M}}|^2_{\text{ann}}(s\longrightarrow t)$, we have that the scattering amplitude is

\begin{equation}
    |\overline{\mathcal{M}}|^2_{\text{sc}}=-\frac{1}{256\pi^4}\lambda^2 Y_\nu^4 m_N^2 t |C_0(0,t,0,m_N^2,m_\Phi^2,m_\Phi^2)|^2.
    \label{eq:amplitude1sc}
\end{equation}

Equivalently, in terms of the scalar and pseudo-scalar mediators, we have

\begin{equation}
    |\overline{\mathcal{M}}|^2_{\text{sc}} = - \frac{\lambda^2 m_N^2 t Y_\nu^4}{1024 \pi ^4} \left(C_0(\phi_1)+C_0(\phi_2)\right)\left(C_0^*(\phi_1)+C_0^*(\phi_2)\right)
    \label{eq:amplitude2sc}
\end{equation}
where the $C_0(\phi_i)$ functions are in terms of the Mandelstam variable $t$.

Furthermore, it is important to perform a thermal average over the scattering cross-section just as we did for the annihilation case. The general definition is

\begin{equation}
\left\langle\sigma_{i j} v_{i j}\right\rangle=\frac{\int d^{3} \mathbf{p}_{i} d^{3} \mathbf{p}_{j} f_{i} f_{j} \sigma_{i j} v_{i j}}{\int d^{3} \mathbf{p}_{i} d^{3} \mathbf{p}_{j} f_{i} f_{j}}.
\label{eq:thermalave}
\end{equation}

We consider the massless neutrino scenario, so dark matter is much heavier than neutrinos and is at rest. The cross-sections in the lab and center-of-mass frames are the same in this approximation. Since the scattering cross-section is only a function of the neutrino energy, the thermal average (Eq. \ref{eq:thermalave}) reduces to

\begin{equation}
\left\langle\sigma_{\nu-\eta}\right\rangle=\frac{\int d^3 \mathbf{p}_{\nu} f_{\nu} \sigma_{\nu-\eta}}{\int d^3 \mathbf{p}_{\nu}f_{\nu}}
\end{equation}

which ends up being a function of neutrino temperature, where 

\begin{equation}
    f_\nu(p_\nu) = \frac{1}{e^{p_\nu/T_\nu}+1}
\end{equation}
is the Fermi-Dirac distribution for the neutrinos. Numerically, the thermally-averaged elastic scattering cross-section is calculated as:

\begin{equation}
\left\langle\sigma_{\nu-\eta}\right\rangle(T_\nu) =
\frac{\int_{0}^{\infty} x^2 \left(\frac{1}{e^{x}+1}\right) \sigma_{\nu-\eta}(xT_{\nu}) dx}
{\int_{0}^{\infty} x^2 \left(\frac{1}{e^{x}+1}\right) dx}
\end{equation}
where $x=E_\nu/T_\nu$.

\section{Z boson decay width}
\label{app:zdecay}
We calculate the Z boson three-body decay to $N\phi_1\nu$ with the formalism of Dalitz invariants \cite{PDG2018} in the approximation that the neutrino mass is small compared to $\phi_1$ and $N$. Our decay rate is given by 

\begin{equation}
\begin{aligned}
\Gamma_{Z\rightarrow N\phi \nu}(m_\phi, m_N, Y_\nu) &= \begin{cases} 
    \Gamma_{Z\rightarrow N\phi \nu_e}+\Gamma_{Z\rightarrow N\phi \nu_\mu}+\Gamma_{Z\rightarrow N\phi \nu_\tau} & \text{if } m_\phi + m_N < m_Z - m_\tau, \\
    \Gamma_{Z\rightarrow N\phi \nu_e}+\Gamma_{Z\rightarrow N\phi \nu_\mu} & \text{if } m_\phi + m_N < m_Z - m_\mu, \\
    \Gamma_{Z\rightarrow N\phi \nu_e} & \text{if } m_\phi + m_N < m_Z, 
\end{cases} \\
\text{where:} \\
\Gamma_Z & = \frac{1}{(2 \pi)^3} \cdot \frac{1}{32 m_Z^3} \int_{m_N^2}^{(m_Z - m_\phi)^2} \text{Integrand}(m_{12}^2, \alpha, \beta, Y_\nu) \, \mathrm{d}m_{12}^2,
\end{aligned}
\end{equation}

and the integrand is: 

\begin{equation}
\begin{aligned}
\text{Integrand}&(m_{12}^2, \alpha, \beta, Y_{\nu}) = \frac{\alpha_l \pi  Y_{\nu}^2}{3 \cos^2\theta_w m_Z^2 \sin^2\theta_w} (\cos^2\theta_w + \sin^2\theta_w)^2 \times \\
& \Bigg\{ 
    m_Z \sqrt{\frac{\left(\frac{m_{12}^2}{m_Z^2} - \beta^2\right)^2}{\frac{m_{12}^2}{m_Z^2}}}
    \sqrt{m_{12}^2 + \frac{m_Z^2 (\alpha^2 - 1)^2}{\frac{m_{12}^2}{m_Z^2}} - 2 m_Z^2 (\alpha^2 + 1)} \ \\
    & \left(
        \frac{m_{12}^4 m_Z^2 (\beta^2 - \alpha^2) +
        \beta^2 m_Z^6 (\beta^2 - \alpha^2 + 1) (\beta^2 + 2) +
        \frac{m_{12}^2}{m_Z^2} m_Z^6 (-2 \beta^4 + 4 \alpha^2 + \beta^2 (2 \alpha^2 - 5))}{\frac{m_{12}^2}{m_Z^2} m_Z^4 (\beta^2 - \alpha^2) - 
        \beta^2 m_Z^4 (\beta^2 - \alpha^2 + 1)}
    \right) \\
    & + 4 m_Z^4 \left(-\frac{m_{12}^2}{m_Z^2} + \alpha^2 + 1\right) 
    \operatorname{ArcCoth}\left(
        \frac{m_Z^2 \left(-\frac{m_{12}^2}{m_Z^2} + \beta^2 + \alpha^2 + 1\right) + 
        \frac{m_Z^4}{m_{12}^2} \beta^2 (1 - \alpha^2)}{
            m_Z \sqrt{\frac{\left(\frac{m_{12}^2}{m_Z^2} - \beta^2\right)^2}{\frac{m_{12}^2}{m_Z^2}}} 
            \sqrt{m_{12}^2 + \frac{m_Z^2 (\alpha^2 - 1)^2}{\frac{m_{12}^2}{m_Z^2}} - 2 m_Z^2 (\alpha^2 + 1)}
        }
    \right)
 \Bigg\}
\end{aligned}
\end{equation}

where $m_{12}^2$ is a Dalitz parameter known as the invariant mass and $\alpha = m_\phi/m_Z$, $\beta = m_N/m_Z$ are the ratios of the masses of the decay products to the mass of the Z boson. 

\begin{figure}[!ht]
    \centering
    \includegraphics[width=0.65\columnwidth]{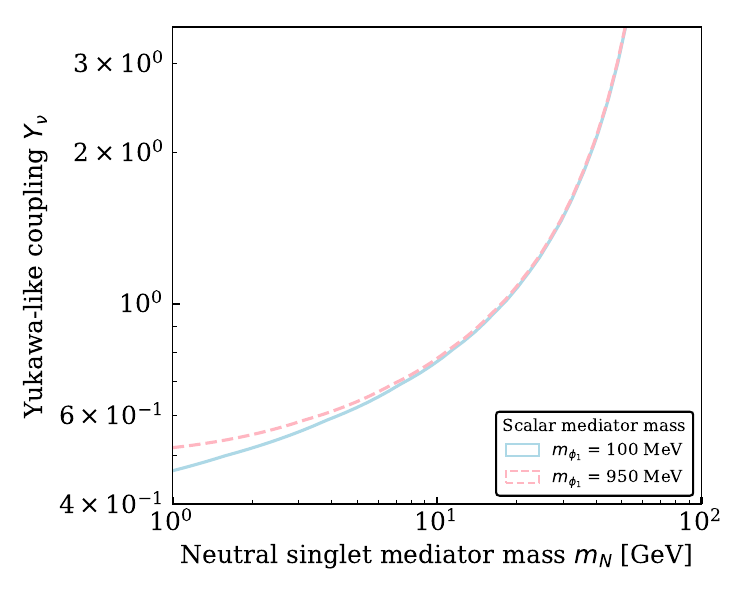}
    \caption{Z boson decay width in the parameter space of the Yukawa-like coupling and the neutral singlet mediator mass for scalar mediator masses of 100 MeV (solid blue line) and 950 MeV (dashed pink line). Both lines represent a decay width of 0.0023 GeV which is the upper limit to the contribution to the total Z decay rate at 90\% C.L.}
    \label{fig:z_decay_constraint_appendix}
\end{figure}

In Fig. \ref{fig:z_decay_constraint_appendix} we note than when the scalar and neutral singlet mediator masses are similar (dashed pink) the decay width constraint is lifted to bigger Yukawa-like couplings by a small amount compared to when the mediators' masses are different (solid blue). At low masses, and when one of the mediators is virtually massless, our calculations agree with those of \cite{Berryman:2018ogk,Brdar:2020nbj}. Our decay width slightly differs at bigger masses due to the inclusion of two massive decay products. Meaning, while in a majoron model as in the case of \cite{Berryman:2018ogk,Brdar:2020nbj} the decay products are $\nu\phi\nu$ and both neutrinos can be taken to be massless, in our model two of the decay products $N\phi\nu$ can be massive and we compute the decay width assuming so. We note that 1-loop contributions may affect these constraints as recently pointed out by \cite{Dev:2024twk}.

To reproduce the results presented in this work visit \cite{ProjectGitHub}.

\end{document}

%% file: jdefs.tex
\def\mnras{MNRAS}%